\documentclass[12pt]{iopart}
\usepackage{epsfig}
\def\({\left(}
\def\){\right)}
\def\[{\left[}
\def\]{\right]}
\def\be{\begin{equation}}
\def\ee{\end{equation}}

\begin{document}
\title{Casimir effect for thin films in QED\\}
\author
{V N  Markov \P \ and  Yu M Pis'mak {\dag \ddag}}
\address
 {\dag \ Department of Theoretical Physics,  State University Saint-Petersburg, Russia}
\address
 {\P \ Department of Theoretical Physics, Saint-Petersburg Nuclear
 Physics Institute, Russia}
\address
  {\ddag \ Institute for Theoretical Physics, University Heidelberg, Germany}
\begin{abstract}
We consider the problem of modeling  of interaction of thin
material films with fields of quantum electrodynamics. Taking into
account the basic principles of quantum electrodynamics (locality,
gauge invariance, renormalizability) we construct a single model
for Casimir-like phenomena arising near the film boundary on
distances much larger then Compton wavelength of the electron. In this region contribution of
Dirac fields fluctuations  are not essential and can be neglected.
In the model the film is presented by a singular background field
concentrated on a 2-dimensional surface and interacting with
quantum electromagnetic field. All properties of the film material
are described by one dimensionless parameter. For two parallel
plane films the Casimir force  appears to be non-universal and
dependent on material property.  It can be both attractive and
repulsive. In the model we study   scattering of electromagnetic
wave on the plane film,  an interaction of plane film with point
charge, homogeneously charged plane  and straight line current.
Here, besides usual results of classical electrodynamics the model
predicts appearance of anomalous electromagnetic phenomena.
\end{abstract}
%\submitto{\JPA}
\maketitle
\section{Introduction}
In 1948 it was shown by Casimir that vacuum fluctuations of
quantum fields generate an attraction between two parallel
uncharged conducting planes \cite{Casimir}. This phenomena called
the Casimir effect (CE) has been well investigated with methods of
modern experiments \cite{Moh1,Moh2, Bressi}. The CE is a
manifestation of influence of fluctuations of quantum fields on
the level of classical interaction of material objects.
Theoretical and experimental investigation of phenomena such a
kind became very important for development of micro-mechanics and
nano-technology.

Though there are many theoretical results on the CE \cite{milton},
however the majority of them are received in framework of several
models based not on the quantum electrodynamics (QED) directly.
Usually, one assumes that the CE can be investigated in the
framework of free massless quantum scalar field theory with fixed
boundary conditions or $\delta$-function potentials
\cite{Jaffe2,Milton2} ignoring restrictions following from gauge
invariance, locality and renormalizability of QED. By means of
such methods one can investigate some of the CE properties, but
there is no possibility of studying other phenomena generated by
interaction of the QED fields with considered classical background
within the same model.

  An approach for construction of the single QED model for investigation
of all peculiar properties of the CE for thin material films was
proposed in \cite{deffect,MarPi}. We consider its
application for simple case of parallel plane films. We show that
gauge invariance, locality and renormalizability  considered as
basic principles make strong restrictions for constructions of the
CE models in QED, which make it possible to reveal new important
features of the CE-like phenomena.

\section{Construction of models}
We construct models for interaction of the material film with  QED
fields on the basis of most general assumptions. We suppose that
the film is presented by a singular background (defect)
concentrated on the 2-dimensional surface. Its interaction with
QED fields has most general form defined by the geometry of the
defect and restrictions following from the basic principles of QED
(gauge invariance, locality, renormalizability). The locality of
interaction means that the action functional of the defect is
represented by an integral over defect surface of the Lagrangian
density which is a polynomial function of space-time point in
respect to fields and derivatives of ones. The coefficients of
this polynomial are the parameters defining defect properties. For
the quantum field theory (QFT) with singular background the
requirement of renormalizibility  was analyzed by Symanzik in
\cite{Sim}. He showed that in order to keep renormalizability of
the model, one needs to add a defect action to the usual bulk
action of QFT model. The defect action must contain all possible
terms with nonnegative dimensions of parameters and not include
any parameters with negative dimensions. In case of QED the defect
action must be also gauge invariant.

     From these requirements it follows  that
for thin film (without charges and currents) which shape is
defined by equation $\Phi ( x ) = 0$, $x=(x_0,x_1,x_2,x_3)$, the
action describing its interaction with photon field $A_\mu(x)$
reads
\begin{equation}
S_{\Phi}(A)= \frac{a}{2} \int
    \varepsilon^{\lambda \mu \nu \rho}
\partial_{\lambda} \Phi ( x ) A_{\mu} ( x ) F_{\nu
\rho} ( x ) \delta ( \Phi ( x))dx \label{v1}
\end{equation}
where $F_{\nu \rho} ( x ) = \partial_{\nu} A_{\rho} -
\partial_{\rho} A_{\nu}$,
$\varepsilon^{\lambda \mu \nu \rho} $ denotes totally
antisymmetric tensor ($\varepsilon^{0123}=1$), $a$ is a constant
dimensionless parameter. The action (\ref{v1}) is a surface Chern-Simon action \cite{Chern,Jackiw}.
The fermion defect action can be written as
\begin{equation}
S_{\Phi}(\bar{\psi},\psi)=  \int \bar{\psi}( x
)[\lambda+u^\mu\gamma_\mu+
\gamma_5(\tau+v^\mu\gamma_\mu)+\omega^{\mu\nu}\sigma_{\mu\nu}]\psi(
x ) \delta ( \Phi ( x))dx \label{v1a}
\end{equation}
Here, $\gamma_\mu$, $\mu=0,1,2,3$, are the Dirac matrices,
$\gamma_5=i\gamma_0\gamma_1\gamma_3\gamma_3$,
$\sigma_{\mu\nu}=i(\gamma_\mu\gamma_\nu-\gamma_\nu\gamma_\mu)/2$,
and $\lambda$, $\tau$, $u_\mu$,$v_\mu$,
$\omega^{\mu\nu}=-\omega^{\nu\mu}$, $\mu, \nu=0,1,2,3$ are 16
dimensionless parameters.

Expressions (\ref{v1}), (\ref{v1a}) are the most general forms of
gauge invariant actions concentrated on the defect surface being
invariant in respect to reparametrization of one and not having
any parameters with negative dimensions.

We consider  in this paper CE-like phenomena arising on the
distances from the defect boundary  much larger then Compton wavelength of the electron.
In this case one can neglect the Dirac fields in QED because of exponential
damping of fluctuations of those on much smaller distances
($\sim m_e^{-1}\approx 10^{-10}cm$ for electron, $\sim m_p^{-1}\approx 10^{-13}cm$ for proton
\cite{deffect}). Thus, for  constructing  of model we can use the action of free
quantum electromagnetic field (photodynamic) with additional
defect action (\ref{v1}).

 For description of all physical phenomena it is enough to calculate
generating functional of Greens functions. For gauge condition
$\phi(A)=0$  it reads
\begin{equation}
G(J)=C\int e^{i S(A,\Phi)+iJA}\delta({\phi(A)})DA \label{va}
\end{equation}
where
\begin{equation}
S(A,\Phi)= -\frac{1}{4}F_{\mu\nu}F^{\mu\nu}+ S_{\Phi}(A),
\label{ac}
\end{equation}
and the constant $C$ is defined by normalization condition
$G(0)|_{a=0}=1$.  The first term on the right hand side of
(\ref{ac}) is the usual action of photon field. Along with defect
action it forms a quadratic in photon field full action of the
system which can be written as $ S(A,\Phi)= 1/2 \ A_\mu
K_\Phi^{\mu\nu}A_{\nu} $.
   The integral (\ref{va}) is gaussian and is
calculated exactly:
$$
 G(J)=
 \mbox{exp}
\left\{ \frac{1}{2}Tr \ln(D_\Phi D^{-1})- \frac{1}{2}JD_\phi J
\right\}
$$
where $D_\Phi$ is the propagator $D_\Phi=iK_\Phi^{-1}$ of
photodynamic with defect in gauge $\phi (A)=0$, and $D$ is the
propagator of photon field without defect in the same gauge. For
the static defect, function $\Phi(x)$ is time independent, and
$\ln G(0)$ defines the Casimir energy.

In order to expose essential Feature of CE-like phenomena in
constructed model, we calculate the Casimir force (CF) for simple
case of two parallel infinite plane films and study a scattering
of electromagnetic wave on the plane defect. We consider also an
interaction of the plane film with a parallel to it straight line
current and an interaction of film with a point charge and
homogeneous charge distribution on parallel plane.

 \section{Casimir force}

We consider defect concentrated on two parallel planes $x_3=0$ and
$x_3=r$. For this model, it is convenient to use a notation like
$x=(x_0,x_1,x_2,x_3)=(\vec{x}, x_3)$. Defect action (\ref{v1}) has
the form:
$$
 S_{2P}=\frac{1}{2}\int(a_1\delta(x_3)+a_2
\delta(x_3-\mbox{r}))
 \varepsilon^{3 \mu\nu \rho} A_{\mu}(x) F_{\nu
\rho}(x)dx.
$$
 The defect action $S_{2P}$  was discussed in \cite{bordag2}
in substantiation  of Chern-Simon type boundary conditions chosen
for studies of the Casimir effect in photodynamics.
This based on boundary conditions approach  is not related
directly to the one we present.  The defect action (\ref{v1}) is
the main point in our model formulation, and no any boundary
conditions are used. The action $S_{2P}$ is translationally
invariant with respect to coordinates $x_i$, $i=0,1,2$. The
propagator $ D_\Phi(x,y)$ is written as:
$$
D_{2P}(x,y) =\frac{1}{(2\pi)^3}\int
D_{2P}(\vec{k},x_3,y_3)e^{i\vec{k}(\vec{x}-\vec{y})}d\vec{k},
$$
and $D_{2P}(\vec{k},x_3,y_3)$ can be calculated exactly. Using
latin indexes for the components of 4-tensors with numbers $0,1,2$
and notations
$$
P^{lm}(\vec{k})=g^{lm}- k^l k^m/{\vec k}^2, \
L^{lm}(\vec{k})=\epsilon^{lmn3}k_n/|\vec{k}|,\
\vec{k}^2=k_0^2-k_1^2-k_2^2, \ |\vec{k}|=\sqrt{\vec{k}^2}
$$
($g$ is metrics tensor), one can present the results for the
Coulomb-like gauge $\partial_0A^0+\partial_1 A^1+\partial_2 A^2=0$
as follows \cite{MarPi}
\begin{eqnarray}
D_{2P}^{33}(\vec{k},x_3,y_3)=\frac{-i \delta(x_3-y_3)}{ |{\vec
k}|^2}, \ D_{2P}^{l3}(\vec{k},x_3,y_3)=
D_{2P}^{3m}(\vec{k},x_3,y_3)=0
 , \nonumber
\\
D_{2P}^{lm}(\vec{k},x_3,y_3) = \frac{P^{lm}(\vec{k}){\cal
P}_1(\vec{k},x_3,y_3)+L^{lm}(\vec{k}){\cal
P}_2(\vec{k},x_3,y_3)}{2| \vec{k} |[( 1 + a_1 a_2 ( e^{2 i |
\vec{k} | \mbox{r}} - 1) )^2 + ( a_1 + a_2 )^2]}  \nonumber
\end{eqnarray}
where
\begin{eqnarray}
{\cal P}_1(\vec{k},x_3,y_3)=
  [ a_1 a_2 - a^2_1 a_2^2 ( 1 - e^{2 i|\vec{k}|
\mbox{r}} ) ]
 [ e^{i| \vec{k} | ( |x_3 | + |y_3 - \mbox{r} | )} +
e^{i| \vec{k} | (
   |x_3 - \mbox{r} | + |y_3 | )} ]  e^{i| \vec{k} |
   \mbox{r}}
   +
   \nonumber
   \\
   +
    [ a^2_1 + a^2_1 a_2^2 ( 1 - e^{2
   i| \vec{k} | \mbox{r}} ) ] e^{i | \vec{k} | ( |x_3
| + |y_3 | )} +
   [ a^2_2 + a^2_1 a_2^2 ( 1 - e^{2 i| \vec{k} |
\mbox{r}} )
   ] e^{i| \vec{k} | ( |x_3 - \mbox{
   r} | + |y_3 -
\mbox{r} | )}
    -
\nonumber
\\
-e^{i| \vec{k} | |x_3 - y_3 |}[( 1 + a_1 a_2 ( e^{2 i | \vec{k} |
\mbox{r}} - 1) )^2 + ( a_1 + a_2 )^2], \nonumber
\\
{\cal P}_2(\vec{k},x_3,y_3)
  =
a_1 [ 1 + a_2 ( a_2 + a_1 e^{2 i| \vec{k} |
   \mbox{r}} ) ]  e^{i | \vec{k} | (
|x_3 | + |y_3 | )} +\nonumber
\\
+ a_2 [
   1 + a_1 ( a_1 + a_2 e^{2 i| \vec{k} | \mbox{r}} )]
e^{i| \vec{k} | ( |x_3 -
   \mbox{r} | + |y_3 - \mbox{r} | )}-
 \nonumber
\\
  -  a_1 a_2 ( a_1 +
   a_2 ) \left( e^{i| \vec{k} | ( |x_3 | + |y_3 -
\mbox{r} | )} + e^{i|
   \vec{k} | ( |x_3 - \mbox{r} | + |y_3 | )} \right)
e^{i| \vec{k} | \mbox{r}}.
   \nonumber
\end{eqnarray}
The energy density $E_{2P}$ of defect is defined as
$$
   \ln G(0)=\frac{1}{2} \ \mbox{Tr} \ln(D_{2P} D^{-1})=-iTS E_{2P}
$$
where $ T=\int dx_0 $ is duration of defect, and $S=\int dx_1 dx_2$,
is the area of film. It is expressed in an explicit form in terms of
polylogarithm function $\mbox{Li}_4(x)$ \cite{MarPi}. For identical
films with $a_1=a_2=a$ it holds: $ E_{2P}= 2E_s+E_{Cas}, E_s =
\int\ln\sqrt{(1+a^2)}\frac{d\vec{k}}{(2\pi)^{3}} , $
\begin{eqnarray}
E_{Cas}=-\frac{1}{16 \pi^2
\mbox{r}^3}\left\{\mbox{Li}_4\left(\frac{a^2}{(a+ i)^2}\right)
+\mbox{Li}_4\left(\frac{a^2}{(a- i)^2}\right) \right\}.\nonumber
\end{eqnarray}
Here $E_s$ is an infinite constant, which can be interpreted as
self-energy density on the plane, and $E_{Cas}$ is an energy
density of their interaction.
 Function $\mbox{Li}_4(x)$ is defined as
$
\mbox{Li}_4(x)=\sum_{k=1}^\infty \frac{x^k}{k^4}=-\frac{1}{2}\int_0
^\infty k^2 \ln(1-xe^{-k})dk.
$
The force $F_{2P}(\mbox{r},a)$ between planes is given by
$$
F_{2P}(\mbox{r},a)=- \frac{\partial E_{Cas}(\mbox{r},a)}{\partial
\mbox{r}}=- \frac{\pi^2}{240 \mbox{r}^4} f(a).
$$
The force $F_{2P}$ is repulsive for $|a|< a_0$ and attractive for
$|a|>a_0$, $a_0\approx 1.03246$ (see Figure 1). For large $|a|$ it
is the same as the usual CF between perfectly conducting planes. The
model predicts that the maximal magnitude of the {\it repulsive}
$F_{2P}$ is expected for $|a|\approx 0.6 $.
\begin{figure}
\centerline{\epsfig{file=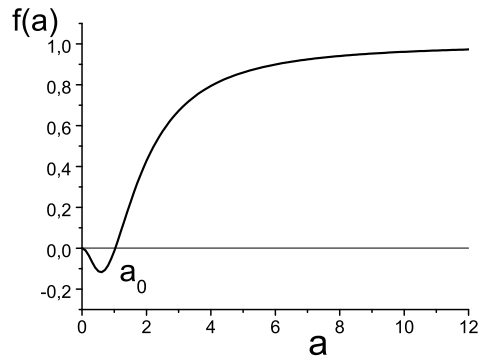,width=7cm}} \caption{Function
$f(a)$ determining Casimir force between parallel planes}
\end{figure}
For two infinitely thick parallel slabs the repulsive CF was
predicted  also in \cite{kenneth}.

 Real film has a finite width, and the
bulk contributions to the CF for nonperfectly conducting slabs with
widths $h_1$, $h_2$ are proportional to $h_1 h_2$. Therefore it
follows directly from the dimensional analysis that the bulk
correction $F_{bulk}$ to the CF is of the form $F_{bulk} \approx
cF_{Cas}h_1h_2/r^2$ where $F_{Cas}$ is the CF for perfectly
conducting planes and $c$ is a dimensionless constant. This
estimation can be relevant for modern experiments on the CE. For
instance, in \cite{Bressi} there were results obtained for parallel
metallic surfaces where width of layer was about $h \approx 50$ nm
and typical distance $r$ between surfaces was $0.5 \mu m \leq r \leq
3 \mu m$. In that case $3\times 10^{-4}\leq (h/r)^2 \leq 10^{-2}$.
In \cite{Bressi} authors have fitted the CF between chromium films
with  function $C_{Cas}/r^4$. They claim that the value of $C_{Cas}$
coincides with known Casimir result within a $15\%$ accuracy.  It
means that bulk force can be neglected, and only surface effects are
essential. In our model the values $a>4.8$ of defect coupling
parameter $a$ are in good agreement with results of \cite{Bressi}.

    Now we study the scattering of classical electromagnetic
wave on plane defect and effects generated by coupling of plane
film with a given classical 4-current.

\section{Interaction of film with classical current and electromagnetic  waves}

  The scattering problem is described in our
approach by a homogeneous classical equation $K^{\mu\nu}_{2P}
A_\nu=0$ of simplified model with $a_1=a$, $a_2=0$. It has a
solution in the form of a plane wave. If one defines transmission
(reflection) coefficient as a ratio $ K_t =
U_t/U_{in}$, ($K_r = U_r/U_{in}$) of transmitted wave  energy $U_t$
(reflected wave  energy $U_r$)  to  incident wave  energy $U_{in}$ , then direct calculations give
the following result: $ K_t = (1 + a^2 )^{-1}, K_r = a^2(1 +
a^2)^{-1} $. We note the following  features of reflection and
transmission coefficients.  In the limit of infinitely large
defect coupling these coefficients coincide with coefficients for
a perfectly conducting plane. The reflection and transmission
coefficients  do not depend on the incidence angle.

 The classical charge  and the wire with current near defect plane are modeled by
appropriately chosen 4-current $J$ in (\ref{va}). The mean vector
potential $ {\cal A}_\mu $ generated by $J$ and the plane $x_3=0$,
with $a_1=a$ can be calculated as
\begin{equation}
 {\cal
A}^\mu=-i\frac{\delta G(J)}{\delta J_\mu}\Bigg|_{a_1=a,a_2=0}=
iD_{2P}^{\mu\nu}J_{\nu}|_{a_1=a,a_2=0}. \label{vecpot}
\end{equation}
Using notations ${\cal F}_{ik}=\partial_i {\cal A}_k-\partial_k
{\cal A}_i$, one can present electric and magnetic fields as ${\vec
E}=({\cal F}_{01},{\cal F}_{02},{\cal F}_{03})$, ${\vec H}=({\cal
F}_{23},{\cal F}_{31},{\cal F}_{12})$. For charge $e$ at the point
$(x_1,x_2,x_3)=(0,0,l)$, $l>0$ the corresponding classical 4-current
is
$$
J_{\mu}(x)=4  \pi e \delta{(x_1)} \delta{(x_2)}
\delta{(x_3-l)}\delta_{0\mu}
$$
In virtue of (\ref{vecpot}) the mean vector potential ${\cal
A}^\mu(x)$ is independent on $x_0$ and  the electric field in
considered system is defined by potential
$$
{\cal A}_0(x_1,x_2,x_3)=\frac{e}{\rho_-}
-\frac{a^2}{a^2+1}\frac{e}{\rho_+}.
$$
where $\rho_+\equiv\sqrt{x_1^2+x_2^2+(|x_3|+l)^2}$,
$\rho_-\equiv\sqrt{x_1^2+x_2^2+(x_3-l)^2}$. The electric field
$\vec{E}=(E_1,E_2,E_3)$ is of the form
$$
E_1=\frac{ex_1}{\rho_-^ {3}}
-\frac{a^2}{a^2+1}\frac{ex_1}{\rho_+^{3}},\
E_2=\frac{ex_2}{\rho_-^ {3}}
-\frac{a^2}{a^2+1}\frac{ex_2}{\rho_+^{3}},\
E_3=\frac{e(x_3-l)}{\rho_-^ {3}}
-\frac{a^2\epsilon(x_3)}{a^2+1}\frac{e(|x_3|+l)}{\rho_+^{3}}.
$$
Here, $\epsilon(x_3)\equiv x_3/|x_3|$.
 We see that for $x_3>0$ the field $\vec{E}$ coincides with field
generated in usual classical electrostatic by charge $e$ placed on
distance $l$ from infinitely thick slab with dielectric constant
$\epsilon=2a^2+1$.

Because ${\cal A}^\mu(x)\neq 0$ for $\mu=1,2,3$, the defect
generate also a magnetic field ${\vec H}=(H_1, H_2, H_3)$:
$$
H_1=\frac{e a x_1}{(a^2+1)\rho_+^3}, \ H_2=\frac{e a
x_2}{(a^2+1)\rho_+^3},\ H_3=\frac{e a(|x_3|+l)}{(a^2+1)\rho_+^3}.
$$
It is an anomalous field which doesn't arise in classical
electrostatics. Its direction depends on sign of $a$.
  In similar one can calculate the fields generated by interaction of the film
and charged plane $x_3=l$, presented by the classical current
$$
J_{\mu}(x)=4  \pi \sigma \delta{(x_3-l)}\delta_{0\mu}.
$$
Here $\sigma$ is the charge density. In this case it holds:
$$
E_1= E_2=0=H_1=H_2=0, \ E_3=
2\pi\sigma\left(\epsilon(x_3-l)-\epsilon(x_3)\frac{a^2}{a^2+1}\right),
\ H_3= 2\pi\sigma\frac{a}{a^2+1}.
$$
Thus, in considered system there is only one dependent on $l$
component of fields $\vec{E}$, $\vec{H}$. It is $E_3$. For
$l\rightarrow\mp\infty$
$$
E_3= 2\pi\sigma\left(\pm 1-\frac{\epsilon(x_3)a^2}{a^2+1}\right),
$$
and for  $l=0$
$$
E_3= \frac{2\pi\sigma\epsilon(x_3)}{a^2+1}.
$$
It is important, to  note that   anomalous fields arise
because the space parity is broken  by  the action (\ref{ac}),
and they are generated in (5) by the $L {\cal P}_2$- term of  propagator $D_{2P}$.

  A current with density $j$ flowing in the wire along the $x_1$-axis is
modeled by
$$
J_{\mu}(x)=4 \pi j  \delta(x_3-l) \delta(x_2) \delta_{\mu 1}
\label{vb10}
$$
For magnetic field from (\ref{vecpot}) one obtains in region
$x_3>0$ the usual results of classical electrodynamics for the
current parallel to infinitely thick slab with permeability
$\mu=(2a^2+1)^{-1}$. There is also an anomalous electric field
$\vec{E}=(0, E_2, E_3)$:
$$
E_2=\frac{2j a}{a^2+1} \frac{x_2}{\tau^2}, \ E_3=\frac{2j a}{a^2+1}
\frac{|x_3|+l}{\tau^2}
$$
where $\tau=(x_2^2+(|x_3|+ l)^2)^{\frac 12}$. Comparing  formulae
$\epsilon=2a^2+1$ and $\mu=(2a^2+1)^{-1}$ for parameter $a$ we
obtain the relation $\epsilon \, \mu=1$. It holds for material of
thick slab interaction of which with point charge and current in
classical electrodynamics was compared with results for thin film of
our model. The speed of light in this hypothetical  material is
equal to one in the vacuum. From the physical point of view, it
could be expected, because interaction of film with photon field is
a surface effect which can not generate the bulk phenomena like
decreasing the speed of light in the considered slab. With this
arguments it seems to be not surprisingly that the reflection
coefficient of electromagnetic wave in our model is independent from
the incidence angle, since by $\epsilon\, \mu=1$ it holds for
Fresnel formulas too.

   The relation $\epsilon \, \mu=1$ is not new in the context of the
Casimir theory. It was first introduced by Brevik and Kolbenstvedt
\cite{Brevik82} who calculated the Casimir surface force density
on the sphere. Only on this condition a contact term turn out to
be zero \cite{Brevik82}. It has been investigated in a number of
subsequent papers. In our approach this condition arises naturally
because we have only one parameter $a$ that must describe both
magnetostatic and electrostatic properties of the film.

The essential property of interaction of films with classical
charge and current is the appearance of anomalous fields. This
fields are suppressed in respect of usual ones by factor $a^{-1}$
and they vanish in case of perfectly conducting plane.
Magnetoelectric (ME)  films are good candidates to detect
anomalous fields and non ideal CE. The generic example of ME
crystals is $Cr_2O_3$ \cite{fiebig}.
     It is important to note that  for ME films
the Lifshitz theory of CE is not relevant but they can be studied
in our approach.

\section{Conclusion}
   The main results of our study on the CE for
thin films in the QED are the following. We have shown that if the
CF holds true for thin material film, then an interaction of
this film with the QED fields can be modeled by photodynamic with
the defect action (\ref{v1}) obtained by most general assumptions
consistent with locality, gauge invariance and renormalizability
of model. Thus, basic principles of QED were essential in our
studies of the CE. These principles make it possible to expose new
peculiarities of the physics of macroscopic objects in QED and
must be taken into account for construction of the models. For
plane films we have demonstrated that the CF is not universal and
depends on properties of the material represented by the parameter
$a$. For $a\rightarrow\infty$ one can obtain the  CF for ideal
conducting planes. In this case the model coincides with
photodynamic considered in \cite{bordag} with boundary condition
$\epsilon^{ijk3}F_{jk}=0$ ($i=0,1,2$) on orthogonal to the
$x_3$-axis planes. For sufficiently small $a$ the CF  appears to
be repulsive. Interaction of plane films with charges and currents
generate anomalous magnetic and electric fields which do not arise
in  classical electrodynamics. The ME materials could be used  for
observation of  phenomena predicted by our model.
 We hope that the obtained theoretical results can be proven
by modern experimental methods.

\section*{Acknowlegements}
We thank  M. Bordag, D.V.Vassilevich, A.N.Vassiliev and F.J.Wegner
for valuable discussions. The work was supported  by Grant
05-02-17477 (V.N. Markov) and Grant 03-01-00837 (Yu.M. Pis'mak)
from Russian Foundation for Basic Research.

\end{document}